\documentclass[10pt,preprint,letterpaper]{article}

\usepackage{amsfonts}
\usepackage{amsmath}
\usepackage{amssymb}
\usepackage{graphicx}
\usepackage[numbers,sort&compress]{natbib}
\title{Greatly enhanced light emission of MoS$_{2}$ using photonic crystal heterojunction}
\author{\small Liu Jiang-tao$^{1,2,3*}$, Tong Hong$^1$, Wu Zhen-Hua$^4$, Huang Jin-Bao$^1$,  Zhou Yun-Song$^{5\dag}$\\  \footnotesize
$^{1}$College of  Mechanical and electrical engineering, Guizhou Minzu University, Guiyang 550025, China \\
\footnotesize  $^{2}$Department of Physics, Nanchang University, Nanchang 330031, Chin\\
 \footnotesize $^{3}$
Institute for Advanced Study,  Nanchang University, Nanchang 330031, China\\
 \footnotesize $^{4}$
Key Laboratory of Microelectronic Devices and Integrated Technology, Institute \\ \footnotesize of Microelec-
tronics, Chinese Academy of Sciences, Beijing 100029, China\\
 \footnotesize $^{5}$
 \footnotesize Department of Physics, Capital Normal University, Beijing 100037, China\\
 \footnotesize   $^{*}$Email: jtliu@semi.ac.cn
  \footnotesize  $^{\dag}$ Email: 263zys@263.net
 }

\begin{document}

\maketitle

\begin{abstract}
We study the effect of one-dimensional (1D) photonic crystal heterojunction (h-PhC) on the light absorption and light emission of monolayer molybdenum disulfide (MoS$_{2}$),  and obtained the analytical solution of the light absorption and emission of two-dimensional materials in 1D h-PhC. Simultaneously enhancing the light absorption and emission of the medium in multiple frequency ranges is easy as h-PhC has more models of photon localization than the common photonic crystal. Result shows that h-PhC can simultaneously enhance the light absorption and emission of MoS$_{2}$ and enhance the photoluminescence spectrum of MoS$_{2}$ by 2-3 orders of magnitude.
\end{abstract}


Two-dimensional (2D) transition metal dichalcogenides (TMDCs), such as MoS$_{2}$ and WSe$_{2}$, are direct-gap semiconductor 2D materials with excellent optical properties and are thus considered the best materials for future optoelectronic devices\cite{NN12QHW,JS16XC,PRL10KFM,NL10AS,NN13OLS,LSA14CFG,S17WF}. The light absorption and emission of 2D TMDCs per unit mass are much higher than that of traditional semiconductor materials. 2D TMDCs typically have a thickness of less than 1 nm, and their light absorption and emission are weak, thus  limiting their application in optoelectronic devices. However, benefit from thin thickness of 2D materials, 2D TMDCs can be combined with optical microstructures, such as photonic crystals, microcavities, and surface plasmas, and then enhance their light absorption\cite{NL15DHL,NL15SB,PRL12ST,PRB12AF,PRB12AFb,APL15WW,NL12MF,NC12ME,NL14SS,PRB08ZZZ,OL13MAV,SC16JGZ,PCCP16XS,SR16YBW,AP14JZ,JQSRT17LL,JAP15CXZ,
AOP16SZ,N17WM,AN10HYJ,OE17HL,CMS16ZY,OM16NA}  and emission\cite{NL15DHL,NL15SB,NC16ZW,SR15KCJL,APL14AS,AM16WG,SC16HC,SC16JL,NL16TG,2DM16CJ,N16YZ,AP16YJN} due to the optical localization in these structures.  Lien et al. \cite{NL15DHL} and Serkan et al. \cite{NL15SB}used surface plasmas or optical multilayers to enhance the light absorption and emission of MoS$_{2}$ or WSe$_{2}$, thus enhancing the photoluminescence (PL) of MoS$_{2}$ or WSe$_{2}$ by 10-30 times.

To further enhance the light emission and absorption of 2D TMDCs, we investigated the effect of photonic crystal heterojunction (h-PhC) on the light absorption and emission of MoS$_{2}$. Similar to the semiconductor heterojunction, h-PhC comprises photonic crystals (PhC) with different lattice constants or shapes\cite{PRB01LLL}. Earlier studies have found that h-PhC that comprise different PhCs can obtain strong light localization in several frequency ranges\cite{PRB01LLL,JPCM03YSZ}. On the basis of these findings, one can place h-PhC that are formed by different PhCs at intervals to form a multimode high-speed optical waveguide.

Thus, if 2D TMDCs are combined with h-PhC, the strong light localization of h-PhC in multiple frequency ranges can simultaneously enhance the light emission and absorption of 2D TMDCs. We therefore conducted a detailed study. h-PhC consists of 1D PhCs with two kinds of crystal lattices that form an h-PhC microcavity structure. To thoroughly understand the light absorption and emission in h-PhC, we first identified the analytical solution of the light absorption and emission of MoS$_{2}$ in h-PhC. The findings indicate that h-PhC can enhance the light absorption and emission of MoS$_{2}$ and enhance the PL spectrum of MoS$_{2}$ by 2-3 orders of magnitude, which has a promising prospect and important application value in fluorescent probe, 2D LED, etc. The analytical solution can be used not only for the light absorption and emission in h-PhC but also for the calculation of other 1D PhC-2D materials composite structures.

\section*{Model and theory}

\begin{figure}
\includegraphics[width=8cm,clip]{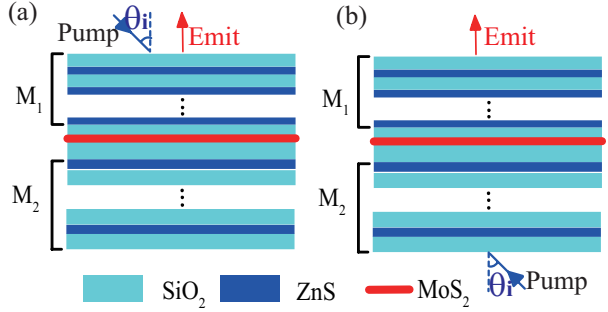}
\caption{Schematic of h-PhC structure. (a) pump light and outgoing light on the same side; (b) pump light and outgoing light on different sides.
}
\label{fig1}%
\end{figure}

    The structure of h-PhC is shown in Fig. 1, i.e., $(A_{1}B_{1})^{N_{1}}C_{1}MC_{2}(B_{2}A_{2})^{N_{2}}$ structure.
$(A_{1}B_{1})^{N_{1}}$ and $(B_{2}A_{2})^{N_{2}}$ layers constitute the two distributed Bragg reflectors (DBRs) , and  $N_{1}$ and $N_{2}$  are the numbers of cycles. The $A_{1}$   and $A_{2}$  layers are made of SiO$_{2}$, and the refractive index  $n_{SiO_{2}}=1.4923+0.81996\lambda'^{2}/(\lambda'^{2}-0.10396^{2})-0.01082\lambda'^{2}$ \cite{AP85EDP}. $\lambda'=\lambda\times10^{6}$, $\lambda$  is wavelength of the input light beams, and the thicknesses of the $A_{1}$  and $A_{2}$  layers are  $\lambda_{10}/(4\times1.53)$ and  $\lambda_{20}/(4\times1.53)$, respectively. $\lambda_{10}$ and $\lambda_{20}$ is the is the center  wavelength of the the upside PhC and the bottom PhC, respectively.  $B_{1}$ and $B_{2}$  layers are composed of  ZnS. The refractive index  $n_{ZnS}=8.393+0.14383/(\lambda'^{2}-0.2421^{2})+4430.99/(\lambda'^{2}-36.72^{2})$ \cite{AO86CAK}. The thicknesses of the $B_{1}$  and   $B_{2}$ layers are $\lambda_{20}/(4\times2.4)$   and  $\lambda_{20}/(4\times2.4)$. The $C_{1}$  and $C_{2}$  layers are made of SiO$_{2}$, and their thicknesses are $d_{C_{1}}$  and  $d_{C_{2}}$, respectively. The M layer is the MoS$_{2}$ layer. Its thickness is 0.65 nm.

To model the absorption of MoS$_{2}$ in this structure, the transfer matrix method
is used first \cite{WI83AY,SR16YBW}. In the \emph{l}th layer, the electric
field of the TE mode light beam with incident angle $\theta_{i}$ is given by
\begin{equation}
{\mathbf{E}}_{l}(x,y)=\left[  A_{l}e^{ik_{x}\left(  x-x_{l}\right)  }%
+B_{l}e^{-ik_{x}\left(  x-x_{l}\right)  }\right]  e^{ik_{y}y}{\mathbf{e}_{z}},
\label{TMM:a1}%
\end{equation}
where $k_{l}=k_{lr}+ik_{li}$ is the wave vector of the incident light,
$\mathbf{e}_{z}$ is the unit vectors in the z direction, and $x_{l}$ is the
position of the \emph{l}th layer in the x direction. And the magnetic field of
the TM mode in the \emph{l}th layer is given by
\begin{equation}
\mathbf{H}_{l}(x,y)=\left[  A_{l}e^{ik_{x}\left(  x-x_{l}\right)  }%
+B_{l}e^{ik_{x}\left(  x-x_{l}\right)  }\right]  e^{ik_{y}y}\mathbf{e}_{z},
\end{equation}
The electric (magnetic) fields of TE (TM) mode in the (\emph{l+1})th and
\emph{l}th layer are related by the matrix utilizing the boundary condition
\begin{align}
\binom{A_{l+1}}{B_{l+1}} &  =\left(
\begin{array}
[c]{cc}%
\frac{\gamma_{l}+\gamma_{l+1}}{2\gamma_{l+1}}e^{ik_{x}d_{l}} & \frac
{\gamma_{l+1}-\gamma_{l}}{2\gamma_{l+1}}e^{-ik_{x}d_{l}}\\
\frac{\gamma_{l+1}-\gamma_{l}}{2\gamma_{l+1}}e^{ik_{x}d_{l}} & \frac
{\gamma_{l}+\gamma_{l+1}}{2\gamma_{l+1}}e^{-ik_{x}d_{l}}%
\end{array}
\right)  \binom{A_{l}}{B_{l}}\nonumber\\
&  =T^{l+1\leftarrow l}\binom{A_{l}}{B_{l}},
\end{align}
where $\gamma_{l}=\frac{k_{lx}^{r}+ik_{lx}^{i}}{\mu_{l}(\omega)}$ ($\gamma
_{l}=\frac{k_{lx}^{r}+ik_{lx}^{i}}{\varepsilon_{l}(\omega)}$)for TE (TM) mode,
$\mu_{l}(\omega)$ is the permeability, $\varepsilon_{l}(\omega)=\varepsilon
_{l}^{r}(\omega)+i\varepsilon_{l}^{i}(\omega)$ is the complex dielectric
permittivity, and $d_{l}$ is the thickness of the \emph{l}th layer. Thus, the
fields in the (\emph{l}+1)th layer are related to the incident fields by the
transfer matrix
\begin{equation}
\binom{A_{l+1}}{B_{l+1}}=T^{l+1\leftarrow l}\cdot\cdot\cdot T^{2\leftarrow
1}T^{1\leftarrow0}\binom{A_{0}}{B_{0}}=\left(
\begin{array}
[c]{cc}%
T_{11} & T_{12}\\
T_{21} & T_{22}%
\end{array}
\right)  \binom{A_{0}}{B_{0}}.
\end{equation}

To thoroughly describe the light absorption and emission of MoS$_{2}$ in h-PhC,
improve the computational speed to optimize the structure, and help scholars
who are not familiar with the transfer matrix method for computing, we
obtained the analytical solution of the light absorption and emission of MoS$_{2}$
in h-PhC using the transfer matrix method. Since the transfer matrix of the
electric fields of TE mode and the transfer matrix of magnetic fields of TM
mode have the same form, we only shows the analytical solution of the TE mode.
First, for a N-period PhC in air, the transfer matrix can be write as\cite{PRE96JMB}
\begin{equation}\binom{A_{0}}{B_{0}}=M_{N}\binom{A_{l+1}}{B_{l+1}}=\left[
\begin{array}
[c]{cc}%
1/t_{N} & r_{N}^{\ast}/t_{N}^{\ast}\\
r_{N}/t_{N} & 1/t_{N}^{\ast}%
\end{array}
\right]  \binom{A_{l+1}}{B_{l+1}},\end{equation}
where $\frac{1}{t_{N}}=\frac{1}{t_{0}}\frac{\sin N\beta}{\sin\beta}-\frac
{\sin\left(  N-1\right)  \beta}{\sin\beta}$, $\frac{r_{N}}{t_{N}}=\frac{r_{0}%
}{t_{0}}\frac{\sin N\beta}{\sin\beta},\beta$ is the Bloch phase in each
period, $t_{0}$ and $r_{0}$ is the transmission  amplitude and reflection
amplitude of the each
period \cite{PRE96JMB}. For the upper part PhC in the h-PhC, the right hand side is not air.
By multiplying the transfer matrix of PhC to the C$_{1}$ layer $T^{C\leftarrow
P}(d_{c}=0)$ and the inverse transfer matrix of PhC to the air layer $\left[
T^{air\leftarrow P}(d_{c}=0)\right]  ^{-1},$ we can get the transfer matrix of
the upper part PhC
\begin{align}
\left(  M_{N_{1}}^{\prime}\right)  & =\left\{  M_{N_{1}}T^{C\leftarrow P}%
(d_{c}=0)\left[  T^{air\leftarrow P}(d_{c}=0)\right]  ^{-1}\right\}  ^{-1}\nonumber\\
& =\frac{1}{2}\left(
\begin{array}
[c]{cc}%
\zeta/t_{N_{1}}+\zeta^{\prime}r_{N_{1}}^{\ast}/t_{N_{1}}^{\ast} &
\zeta^{\prime}t_{N_{1}}+\zeta r_{N_{1}}^{\ast}/t_{N_{1}}^{\ast}\\
\zeta r_{N_{1}}/t_{N_{1}}+\zeta^{\prime}/t_{N_{1}}^{\ast} & \zeta^{\prime
}r_{N_{1}}/t_{N_{1}}+\zeta/t_{N_{1}}^{\ast}%
\end{array}
\right)\nonumber \\
& =\left[
\begin{array}
[c]{cc}%
1/t_{N_{1}}^{\prime} & r_{N_{1}}^{\prime\ast}/t_{N_{1}}^{\prime\ast}\\
r_{N_{1}}^{\prime}/t_{N_{1}}^{\prime} & 1/t_{N_{1}}^{\prime\ast}%
\end{array}
\right]  ,\end{align}

where $\zeta=1+\sqrt{\varepsilon_{c}^{{}}}\cos\theta_{c},\zeta^{\prime
}=1-\sqrt{\varepsilon_{c}^{{}}}\cos\theta_{c},\varepsilon_{c}^{{}}%
=\varepsilon_{c_{1}}^{{}}$ $=\varepsilon_{c_{2}}^{{}}$ is the refractive index of C$_{1}$ and C$_{2}$ layers,$\theta_{c}=\theta
_{c_{1}}=\theta_{c_{2}}$ is the propagation angle in the C$_{1}$ and C$_{2}$
layer. Similar, we can get the transfer matrix of the lower part PhC,

\begin{align}M_{N_{2}}^{\prime}&=\left\{  \left[  T^{C\leftarrow P}(d_{c}=0)\right]
^{-1}T^{air\leftarrow E}(d_{c}=0)\right\}  M_{N_{2}}=\nonumber\\
&=\frac{1}{2}\left(
\begin{array}
[c]{cc}%
\zeta/t_{N_{2}}+\zeta^{\prime}r_{N_{2}}/t_{N_{2}} & \zeta^{\prime}/t_{N_{2}%
}^{\ast}+\zeta r_{N_{2}}^{\ast}/t_{N_{2}}^{\ast}\\
\zeta r_{N_{2}}/t_{N_{2}}+\zeta^{\prime}/t_{N_{2}} & \zeta^{\prime}r_{N_{2}%
}^{\ast}/t_{N_{2}}^{\ast}+\zeta/t_{N_{2}}^{\ast}%
\end{array}
\right)  \nonumber\\
&=\left[
\begin{array}
[c]{cc}%
1/t_{N_{2}}^{\prime} & r_{N_{2}}^{\prime\ast}/t_{N_{2}}^{\prime\ast}\\
r_{N_{2}}^{\prime}/t_{N_{2}}^{\prime} & 1/t_{N_{2}}^{\prime\ast}%
\end{array}
\right].  \end{align}

The transfer matrix of the C$_{1}$ layer is \cite{PRB12AF}
\begin{equation}M_{f}\left(  d_{C_{1}}\right)  =\left[
\begin{array}
[c]{cc}%
e^{-ik_{cx}d_{C_{1}}} & 0\\
0 & e^{ik_{cx}d_{C_{1}}}%
\end{array}
\right],\end{equation}
and the transfer matrix of the C$_{2}$ layer is
\begin{equation}M_{f}\left(  L_{cav}-d_{C_{1}}\right)  =\left[
\begin{array}
[c]{cc}%
e^{-ik_{cx}\left(  L_{cav}-d_{C_{1}}\right)  } & 0\\
0 & e^{ik_{cx}\left(  L_{cav}-d_{C_{1}}\right)  }%
\end{array}
\right]  ,\end{equation}

where $L_{cav}=d_{C_{1}}+d_{C_{2}}$ is the microcavity length, $k_{cx}$ wave
vector of the light in the C$_{1}$ or C$_{2}$ layer. Take the approximate
$e^{ik_{Mx}d_{M}}\approx1+ik_{Mx}d_{M}^{{}}$, where $k_{Mx}$ wave vector of
the light in the MoS$_{2}$ layer and $d_{M}^{{}}$ is the thickness of the and MoS$_{2}$
layer, the transfer matrix of the MoS$_{2}$ layer is

\begin{equation}M_{MoS_{2}}=\left[  T^{C\leftarrow MoS_{2}}T^{MoS_{2}\leftarrow
C}(d=0)\right]  ^{-1}=\left[
\begin{array}
[c]{cc}%
1-\eta_{1} & -\eta_{2}\\
\eta_{2} & 1+\eta_{1}%
\end{array}
\right],  \end{equation}
where $\eta_{1}=ik_{Mx}d_{M}^{{}}\left[  \left(  \sqrt{\varepsilon_{MoS_{2}%
}/\varepsilon_{C}}\cos\theta_{MoS_{2}}/\cos\theta_{c}\right)  +\left(
\sqrt{\varepsilon_{c}/\varepsilon_{MoS_{2}}}\cos\theta_{c}/\cos\theta
_{MoS_{2}}\right)  \right]  /2$, and $\eta_{2}=ik_{Mx}d_{M}^{{}}\left[
\left(  \sqrt{\varepsilon_{MoS_{2}}/\varepsilon_{C}}\cos\theta_{MoS_{2}}%
/\cos\theta_{c}\right)  -\left(  \sqrt{\varepsilon_{c}/\varepsilon_{MoS_{2}}%
}\cos\theta_{c}/\cos\theta_{MoS_{2}}\right)  \right]  /2$. Thus, the total transfer matrix of the $C_{1}$, $C_{2}$, and MoS$_{2}$ layer is
\begin{equation}M_{f}\left( d_{C_{1}}\right)  M_{MoS_{2}}M_{f}\left(  L-d_{C_{1}}\right)  =\left[
\begin{array}
[c]{cc}%
\left(  1-\eta_{1}\right)  e^{-ikL} & -\eta_{2}e^{ik\left(  L-2d_{C_{1}}\right)
}\\
\eta_{2}e^{-ik\left(  L-d_{C_{1}}\right)  } & \left(  1+\eta_{1}\right)  e^{ikL}%
\end{array}
\right].\end{equation}
 The total transfer matrix of the h-PhC is
\begin{equation}M=M_{N_{1}}^{\prime}M_{f}\left(d_{C_{1}}\right)  M_{MoS_{2}}M_{f}\left(
L-d_{C_{1}}\right)  M_{N_{2}}^{\prime}.\end{equation}
we can get the matrix element
\begin{equation}M_{11}=\left[  \left(  1-\eta_{1}\right)/\varphi_{1}t_{N_{1}}^{\prime}%
-\eta_{2}\varphi_{2}r_{N_{2}}^{\prime}/t_{N_{1}}^{\prime
}+\eta_{2}r_{N_{1}}^{\prime\ast}/\varphi_{2}t_{N_{1}%
}^{\prime\ast}+\left(  1+\eta_{1}\right)  \varphi_{1}r_{N_{2}}^{\prime}r_{N_{1}%
}^{\prime\ast}/t_{N_{1}}^{\prime\ast}\right]  /t_{N_{2}}^{\prime},\end{equation}
and
\begin{equation}M_{21}=\left[  \left(  1-\eta_{1}\right)r_{N_{1}}^{\prime}%
/\varphi_{1}t_{N_{1}}^{\prime}-\eta_{2}\varphi_{2}r_{N_{2}}^{\prime
}r_{N_{1}}^{\prime}/t_{N_{1}}^{\prime}+\eta_{2}/\varphi_{2}t_{N_{1}}^{\prime\ast}+\left(  1+\eta_{1}\right)  \varphi_{1}r_{N_{2}}^{\prime
}/t_{N_{1}}^{\prime\ast}\right]  /t_{N_{2}}^{\prime},\end{equation}
where $\varphi_{1}= e^{ikL}$, $\varphi_{2}=e^{ik\left(  L-2d_{C_{1}}\right)}$
The transmittance of the h-PhC is $T=\left\vert 1/M_{11}\right\vert ^{2}$;
 the reflectance of the h-PhC is $R=\left\vert M_{21}/M_{11}\right\vert
^{2}$ \cite{PRB12AF,PRE96JMB}; the absorption of MoS$_{2}$ layer $A_{MoS_{2}}=1-R-T$.

The spontaneous emission of the monolayer MoS$_{2}$  in the h-PhC can be
treated as two emitted correlated wavepackets, upward (downward) propagating
wave packet $\mathcal{P}_{u}$ ($\mathcal{P}_{d}$).  The emission wavepackets
are partially transmitted and reflected by the two DBRs. The filed amplitude
of the light emitted out from the exit DBR mirror is given by
\cite{OE11SH,JMO94DGD,EPL15FFY}
\begin{align}
E_{DBRt}\left(  t\right)   &  =t_{t}\mathcal{P}(t)+t_{t}r_{b}\mathcal{P}%
(t-\frac{2z_{ol}}{c})\nonumber\\
&  +t_{t}(r_{b}r_{t})\mathcal{P}(t-\frac{2L_{oc}}{c})\nonumber\\
&  +t_{t}(r_{b}r_{b}r_{t})\mathcal{P}(t-\frac{2z_{ol}}{c}-\frac{2L_{oc}}%
{c})+\cdot\cdot\cdot,
\end{align}
where $r_{t}$ and $t_{t}$ is the reflection amplitude and transmission
amplitude of the exit  DBR mirror, respectively, $r_{b}$ is the reflection
amplitude of the back DBR mirror, $z_{ol}$ is the distance between the
monolayer MoS$_{2}$ and back DBR mirror. When the pump and outgoing lights
are on the same side of the h-PhC, $r_{t}=r_{N_{1}}^{\prime},$ $\ t_{t}%
=t_{N_{1}}^{\prime},$ \ $r_{b}=$\ $r_{N_{2}}^{\prime}$, $z_{ol}=d_{C_{2}}$;
When the pump and outgoing lights are on the different side of the h-PhC,
$r_{t}=r_{N_{2}}^{\prime},$ $\ t_{t}=t_{N_{2}}^{\prime},$ \ $r_{b}%
=$\ $r_{N_{1}}^{\prime}$, $z_{ol}=d_{C_{1}}$;  $\mathcal{P}(t)$ is the
electric amplitude against time for a single emission event (in either
direction), $L_{oc}=n_{c}L_{cav}=n_{c}\left(  d_{C_{1}}+d_{C_{2}}\right)  $ is
the optical length of microcavity, $n_{c}$ is the refractive index of the
C$_{1}$ and C$_{2}$  layer. By using the Fourier transform, the emitted
radiation from the top DBR mirror in the frequency domain can be written as
\begin{align}
E_{DBRt}\left(  \omega\right)   &  =\frac{t_{t}}{2\pi}\int_{-\infty}^{\infty
}\mathcal{P}(t)\exp(i\omega t)dt\nonumber\\
&  +\frac{t_{t}r_{b}}{2\pi}\int_{-\infty}^{\infty}\mathcal{P}(t-\frac{2z_{ol}%
}{c})\exp(i\omega t)dt\nonumber\\
&  +\frac{t_{t}(r_{b}r_{t})}{2\pi}\int_{-\infty}^{\infty}\mathcal{P}%
(t-\frac{2L_{oc}}{c})\exp(i\omega t)dt\nonumber\\
&  +\frac{t_{t}(r_{b}r_{b}r_{t})}{2\pi}\int_{-\infty}^{\infty}\mathcal{P}%
(t-\frac{2z_{ol}}{c}-\frac{2L_{oc}}{c})\nonumber\\
&  \times\exp(i\omega t)dt+\cdot\cdot\cdot.\label{emt:a2}%
\end{align}
Neglected the changes of the spontaneous time, integral of Eq. (\ref{emt:a2}),
the  emission intensity can be calculated by \cite{OE11SH,JMO94DGD}
\begin{align}
|E_{DBRt}\left(  \lambda\right)  |^{2} &  =\frac{1+R_{b}+2\sqrt{R_{b}}%
\cos\left(  \frac{4\pi n_{c}Z_{OL}}{\lambda}\right)  }{1+R_{b}R_{t}%
-2\sqrt{R_{b}R_{t}}\cos\left(  \frac{4\pi n_{c}L_{cav}}{\lambda}\right)
}\nonumber\\
&  \times T_{t}|\mathcal{P}\left(  \lambda\right)  |^{2},
\end{align}
where $T_{t}=|t_{t}|^{2}$ and $R_{t}=|r_{t}|^{2}$ is the transmittance and
reflectance of the exit DBR mirror, respectively, $T_{b}=|t_{b}|^{2}$ and
$R_{b}=|r_{b}|^{2}$ is the the transmittance and reflectance of the back DBR
mirror with the MoS$_{2}$ TFT, respectively.

\section*{RESULTS}

\begin{figure}
\includegraphics[width=8cm,clip]{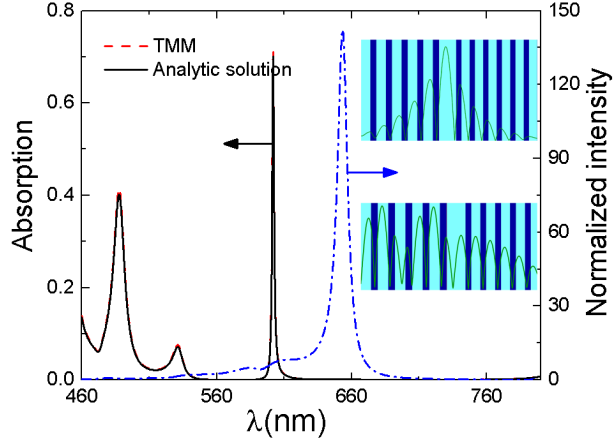}
\caption{The absorption  and relative radiation intensity of MoS$_{2}$ when the pump and outgoing lights are on the same side of the h-PhC. The black solid line and the red dashed line are the calculation results of the analytical solution and the transfer matrix method, respectively. The upper and lower illustrations are the light field distribution at wavelengths of 488  and 602 nm, respectively.
}
\label{fig2}%
\end{figure}

We first calculated the absorption  and the relative radiation intensity of MoS$_{2}$ when the pump and outgoing lights are on the same side of the h-PhC. The calculated parameters are as follows:  $\lambda_{10}=730$ nm, $\lambda_{20}=630$ nm,  $d_{C_{1}}=0$ nm,  $d_{C_{2}}=214$ nm,  $N_{1}=6$, and  $N_{2}=7$. The incident angle  $\theta_{i}=48^{\circ}$. The pump light is in TE mode. The outgoing light is vertically emitted. Two strong absorption peaks emerge at wavelengths of 488 (consistent with the wavelength of the pump light used in the experiment\cite{NL15SB}) and 602 nm. Quite  small difference can be found between the calculation results of the analytical solution and the transfer matrix method  due to  the approximate
$e^{ik_{Mx}d_{M}}\approx1+ik_{Mx}d_{M}^{{}}$ is used.  The optical wavelength of 602 nm is in the bandgap of the two PhCs with strong localization properties (upper illustration of Figure 1) and strong absorption. The absorption  can reach 0.7 or more, which is approximately 6 times more than that without h-PhC. If the wavelength of the pump light is 488 nm, it is only in the bandgap of the bottom PhC. The localization is weak. The absorption  is 0.4, which is about 3 times more than that without h-PhC. The emission efficiency is enhanced by 140 times due to the high reflectivity of the PhCs on both sides. Thus, when the pump lights are 488 and 602 nm, the PL is enhanced by approximately 420 and 840 times, respectively.

\begin{figure}
\includegraphics[width=8cm,clip]{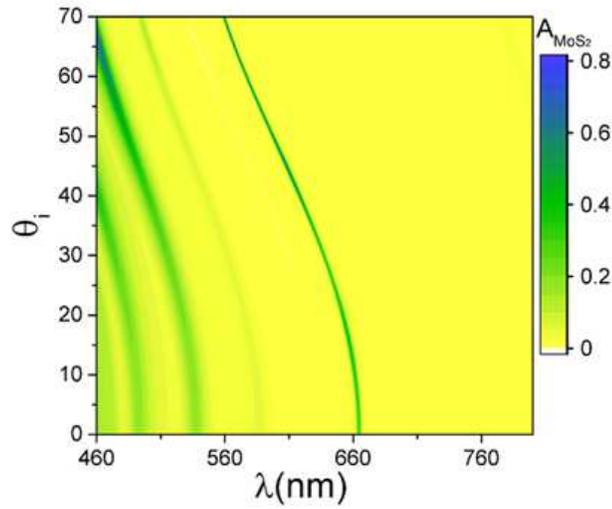}
\caption{h-PhC MoS$_{2}$ absorption changes as the incident angle and wavelength change.
}
\label{fig3}%
\end{figure}

The MoS$_{2}$ absorption  in h-PhC is sensitive to the incidence angle. The calculation results are shown in Figure 3. The resonant wavelength of the microcavity is  $m_{0}\lambda_{c}/2=L_{c}cos\theta'$.   $L_{c}=n_{c}d_{c} $is the microcavity optical path,   $m_{0}$ is a positive integer, and $\theta'=arcsin\theta_{i}/n_{c}$ is the propagation angle of light in the defective layer. Thus, when the incident angle increases, the resonance wavelength moves in the short-wave direction, the reflectivity of the PhCs on both sides increases, the travel path of light in MoS$_{2}$ increases, and the maximum absorption  can reach 0.8 or more. PL is enhanced by approximately 3 orders of magnitude.

\begin{figure}
\includegraphics[width=8cm,clip]{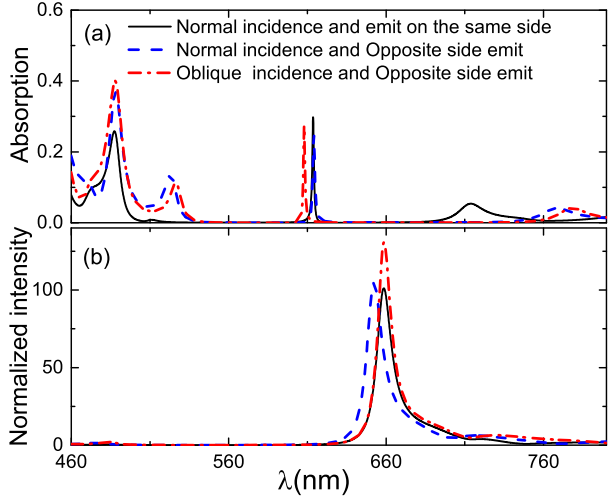}
\caption{(a) Light absorption and (b) light emission of MoS$_{2}$ in h-PhC under different pump lights. The black solid line: when the pump light is normally incident, the pump and outgoing lights are in the same side of the h-PhC; the blue dashed line: when the pump light is normally incident, the pump and outgoing lights are on different sides of the h-PhC; the red dash-dotted line: when the pump light is obliquely incident, the pump and outgoing lights are on different sides of the h-PhC.
}
\label{fig4}%
\end{figure}

We also calculated the positive incidence of the pump light and the absorption  and relative radiation intensity of MoS$_{2}$ when the pump and outgoing lights are on different sides of the h-PhC. As in the experiment,  in our calculation,  the wavelength of the pump light is 488 nm, and the wavelength of the outgoing light approaches 660 nm. We calculated the corresponding parameters by optimizing the h-PhC structure under different pump light incidences as follows: when the pump light is normally incident and the pump and outgoing lights are on the same side of the h-PhC,  $\lambda_{10}=580$ nm, $\lambda_{20}=610$ nm,  $d_{C_{1}}=2$ nm,  $d_{C_{2}}=214$ nm,  $N_{1}=7$, and  $N_{2}=7$. When the pump light is normally incident and the pump and outgoing lights are on different sides of the h-PhC, $\lambda_{10}=630$ nm, $\lambda_{20}=570$ nm,  $d_{C_{1}}=11$ nm,  $d_{C_{2}}=203$ nm,  $N_{1}=7$, and  $N_{2}=7$.  When the pump light is obliquely incident and the pump and outgoing lights are on different sides of the h-PhC,  $\lambda_{10}=660$ nm, $\lambda_{20}=580$ nm,  $d_{C_{1}}=0$ nm,  $d_{C_{2}}=216$ nm,  $N_{1}=7$, $N_{2}=7$,  and  $\theta_{i}=30^{\circ}$.

The detailed calculation results are shown in Figure 4. Regardless of whether the pump and outgoing lights are on the same or different sides of the h-PhC, the absorption  of MoS$_{2}$ is higher when the pump light is obliquely incident. A strong local touch in the vicinity of two wavelengths can be easily obtained as change in incidence angle can adjust the resonant wavelength. When the pump light is normally incident and the pump and outgoing lights are on different sides of the h-PhC, the absorption  of MoS$_{2}$ is high because if MoS$_{2}$ in the microcavity structure obtains strong absorption and emission, the reflectivity of the rear reflector should be higher but the reflectivity of the front reflector should not be excessively high \cite{OL13MAV}. The bandgap width of the PhC is not big enough due to the large difference between the wavelengths of the pump and outgoing lights. The pump and outgoing lights on different sides realize this goal easily.

\begin{figure}
\includegraphics[width=8cm,clip]{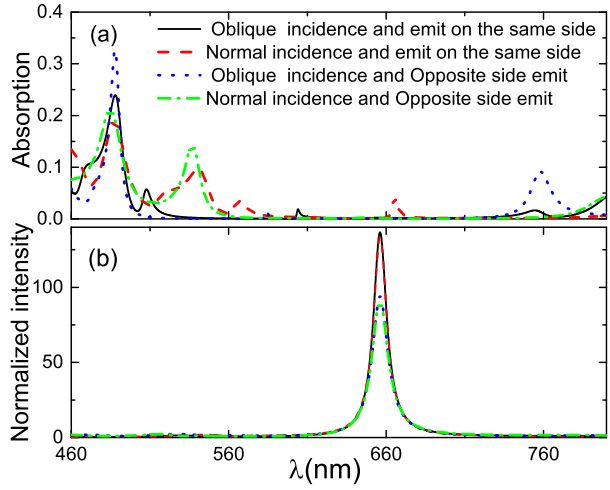}
\caption{(a) Light absorption and (b) light emission of MoS$_{2}$ in homojunction PhC under different pump lights. The black solid line: when the pump light is obliquely incident and the pump and outgoing lights are on the same side; the red dashed line: when the pump light is normally incident and the pump and outgoing lights are on the same side; the blue dotted line: when the pump light is obliquely incident and the pump and outgoing lights are on different sides; the green dash-dotted line: when the pump light is normally incident and the pump and outgoing lights are on different sides.
}
\label{fig5}%
\end{figure}

For comparison, we calculated the light absorption and emission of MoS$_{2}$ in a homojunction. The optimized structural parameters are as follows: when the pump light is obliquely incident and the pump and outgoing lights are on the same side,  $\lambda_{10}=\lambda_{20}=660$ nm, $d_{C_{1}}=0$ nm,  $d_{C_{2}}=215$ nm,  $N_{1}=5$, $N_{2}=7$, and  $\theta_{i}=42^{\circ}$. When the pump light is normally incident and the pump and outgoing lights are on the same side,  $\lambda_{10}=\lambda_{20}=680$ nm, $d_{C_{1}}=4$ nm,  $d_{C_{2}}=210$ nm,  $N_{1}=5$, and $N_{2}=7$. When the pump light is obliquely incident and the pump and outgoing lights are on different sides,  $\lambda_{10}=\lambda_{20}=670$ nm, $d_{C_{1}}=0$ nm,  $d_{C_{2}}=214$ nm,  $N_{1}=6$, $N_{2}=5$, and  $\theta_{i}=54^{\circ}$. When the pump light is normally incident and the pump and outgoing lights are on different sides:  $\lambda_{10}=\lambda_{20}=650$ nm, $d_{C_{1}}=0$ nm,  $d_{C_{2}}=214$ nm,  $N_{1}=6$, and $N_{2}=5$. These structures show that the localization of homojunction PhC is not excessive and increasing the light absorption and emission of MoS$_{2}$ at the same time  is difficult. When the light emission is strong, the light absorption is usually low, with an absorption  of only approximately 0.2. When the pump light is obliquely incident and the pump and outgoing lights are on different sides, the absorption  is the largest (approximately 0.33) but the outgoing light enhancement is low. If light emission is enhanced using longer PhC cycles than those used in the current study, the light absorption of MoS$_{2}$ will decrease. However, this case does not happen in h-PhC.

Finally, we discuss the effect of light localization and the feasibility of the experiment.

The effect of light localization: We used the Q value to judge the strength of light localization. The larger the Q value, the higher the light localization and light absorption and emission intensity of MoS$_{2}$. However, the higher the Q value, the smaller the full width at half maximum of the spectral line and the narrower the absorption and PL spectra. Excessively narrow absorption and emission spectra are not conducive to practical application.  Moreover, when the Q value is high, the microcavity affects the transition time of the exciton. Thus, in Optimizing the parameters, we choose $N_{1}\leq7$ and $N_{2}\leq7$.

The feasibility of the experiment: PhC and 2D materials composite structures (particularly 2D materials-PhC microcavity) were created \cite{NL12MF,NC12ME,NL14SS}. Compared with the existing structure, this structure only changes the lattice constant of the upper and lower reflectors of the PhC microcavity. Therefore, the experiment is completely achievable.

\section*{Conclusion}
We studied the effect of 1D h-PhC on the light absorption and emission of monolayer MoS$_{2}$ and obtained the analytical solution of light absorption and emission in 1D PhC-2D materials composite structures. h-PhC has more models of photon localization than common PhC, enhances the light emission and absorption of MoS$_{2}$ simultaneously, and increases the PL spectrum of MoS$_{2}$ by 2-3 orders of magnitude. When the pump light is obliquely incident and the pump and outgoing lights are on different sides of the h-PhC, it is easier to enhance the light absorption and emission of MoS$_{2}$ can be enhanced at the same time. The analytical solution can be used not only for light absorption and emission in h-PhC but also for the calculation of light emission and absorption of other 1D PhC-2D materials composite structures. This study has a promising prospect and important application value in 2D material-based optoelectronic devices.

\section*{METHODS}
The modified transfer matrix method is used to model the absorption of  monolayer MoS$_{2}$ in the photonic crystal micro-cavity.

\bibliography{mybib}

\section*{Acknowledgements}This work was supported by the NSFC (Grant Nos. 11364033,
11764008, 61774168), Project of master's excellent talent program in guizhou province (ZYRC[2014]008), and the Innovation Group Major Research Project of Department of Educationin in Guizhou Province (No. KY[2016]028)).

\section*{Author contributions}
J.T.L. supervised the project, did the theoretical derivation and
 the numerical calculation, analyzed the results, and wrote the paper.  T. H., W. Z. H., and W. J. B. analyzed of the results and wrote the paper. Z. Y. S. supervised the project, analyzed the results, and wrote the paper. All authors discussed the results and commented on the manuscript.


\emph{Conflict of Interest}: The authors declare no competing
financial interest.


\end{document}